\documentclass[a4paper,11pt]{article}
\usepackage{jheppub} 
\usepackage{booktabs}
\usepackage{subcaption}
\usepackage{indentfirst}

\title{\boldmath Lifshitz transition in a holographic finite density flavour brane Weyl semimetal}

\author{Cheng-Yuan Lu$^{1}$, Xian-Hui Ge$^{1,2*}$, Sang-Jin Sin$^{3}$}
\affiliation{$^1$Department of Physics, Shanghai University,\\
Shanghai 200444,  China}
\affiliation{$^2$Shanghai Key Laboratory of High Temperature Superconductors,\\
Shanghai 200444,  China}
\affiliation{$^3$Department of Physics, Hanyang University,\\
Seoul, 04763, South Korea}
\emailAdd{luchengyuan@shu.edu.cn, gexh@shu.edu.cn$^*$, sjsin@hanyang.ac.kr}

\abstract{We extend a top-down holographic model of a Weyl semimetal to finite charge density and compute the fermionic spectral function by introducing two probe fermions of opposite chirality. The model is controlled by the boundary fermion mass $M$ and the chemical potential $\mu$. In the zero density, small-$M$ limit, we recover four energy bands, two Weyl points, and linear dispersion in their vicinity, the hallmarks of a Weyl semimetal. As $M$ increases, the bands between the Weyl points become progressively compressed and the spectral weight associated with those bands is smeared out. At finite charge density, we map the Fermi surface in momentum space and identify a Lifshitz transition: two distinct Fermi pockets, each enclosing a different Weyl point, merge into a single large Fermi surface that encloses both. This transition can be induced by either control parameter. Varying $M$ alters the band structure and thus the band shape, which drives the Lifshitz transition, whereas changing $\mu$ shifts the bands relative to the Fermi level without qualitatively changing the band structure, producing the Lifshitz transition by moving the band positions.}

\begin{document}
\maketitle
\flushbottom

\section{Introduction}
\label{sec:intro}
Weyl semimetals are materials in which the conduction and valence bands cross at isolated points in momentum space \cite{yan2017topological}. These band-crossing points—called Weyl points—occur in pairs and host low-energy excitations that are described by Weyl fermions of opposite chirality, with a linear dispersion relation near each point. In 1929, H. Weyl introduced the concept of Weyl fermions as fundamental particles \cite{weyl1929electron}. But they have not been observed as elementary particles in the vacuum. Until 2015, people discovered Weyl semimetals TaAs, in which Weyl fermions appear as emergent quasiparticles \cite{lv2015experimental,xu2015discovery}. 

Weyl semimetals represent a class of topological semimetals whose properties are characterized by the Chern number, a topological invariant tied to the chirality of Weyl points. These Weyl points are topologically protected, ensuring the robustness of the associated electronic states against perturbations. A topologically protected Weyl point with fixed chirality behaves analogously to a magnetic monopole in momentum space, with the Chern number arising from the integration of the Berry curvature around it \cite{weng2015quantum}. As a result, Weyl semimetals exhibit an intrinsic anomalous Hall effect and host distinctive surface states known as Fermi arcs \cite{jia2016weyl}.

Weyl physics arises when either time-reversal symmetry $\mathcal{T}$ or inversion symmetry $\mathcal{P}$ is broken, which splits a Dirac point into a pair of Weyl points of opposite chirality. This phenomenon admits a compact description in terms of a minimal fermionic field theory \cite{Colladay:1998fq}
\begin{align}
    \mathcal{L}=\bar{\psi}\left(i \gamma^\mu \partial_\mu-m+B_j^5 \gamma^j \gamma^5\right) \psi,
    \label{Lagrangian}
\end{align}
where $\mu$ labels the spacetime coordinate, $m$ is the fermion mass, $\gamma^{\mu}$ are the Gamma matrices, and $\gamma^{5}=i\gamma^{t}\gamma^{x}\gamma^{y}\gamma^{z}$. The axial coupling $B_j^5$ breaks $\mathcal{T}$. In particular, choosing $B_j^5=\frac{b}{2}\delta_{jz}$ gives the band structure illustrated below.

\begin{figure}[htbp]
    \centering
    \begin{subfigure}{0.48\textwidth}
        \centering
        \includegraphics[width=\textwidth]{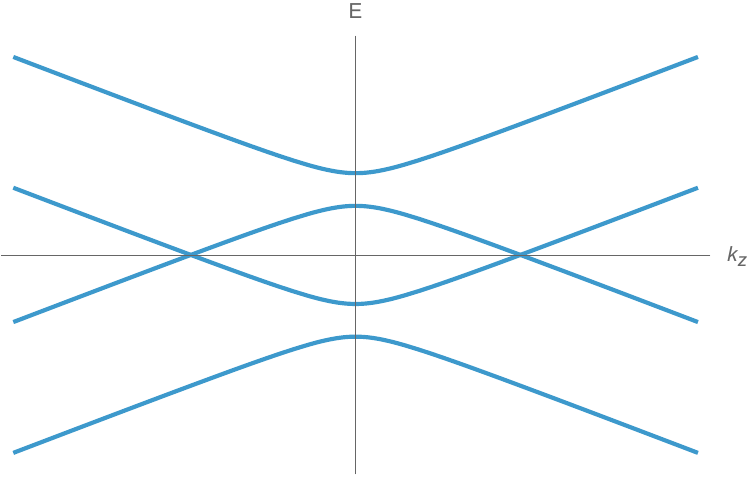}
        \caption{$|\frac{m}{b}|<\frac{1}{2}$}
        \label{fig:1a}
    \end{subfigure}
    \begin{subfigure}{0.48\textwidth}
        \centering
    \includegraphics[width=\textwidth]{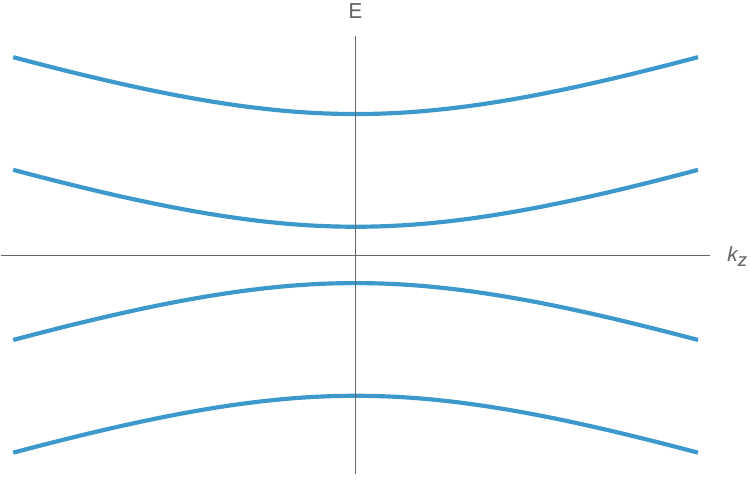}
        \caption{$|\frac{m}{b}|>\frac{1}{2}$}
        \label{fig:1b}
    \end{subfigure}
    \caption{The band structure obtained from the Weyl semimetal Lagrangian \eqref{Lagrangian}. (a) For $|\frac{m}{b}|<\frac{1}{2}$, the two bands intersect on the $k_z$ axis at two Weyl points, each exhibiting a linear dispersion. (b) For $|\frac{m}{b}|>\frac{1}{2}$, the band crossing is lifted and a full gap opens.}
    \label{fig:1}
\end{figure}

The Lagrangian \eqref{Lagrangian} describes the free Dirac fermion theory. TaAs is a weakly correlated Weyl semimetal, and its band structure is well reproduced by density functional theory \cite{xu2015discovery}. As more materials have been discovered, evidence has emerged for strongly correlated Weyl semimetals. In particular, Weyl–Kondo semimetals were theoretically proposed as a strongly correlated route to Weyl physics \cite{Lai2018WeylKondo}, and $\text{Ce}_3\text{Bi}_4\text{Pd}_3$ has been reported experimentally as a Weyl–Kondo candidate \cite{Dzsaber:2021ucs}. However, this interpretation remains controversial: strong Kondo/ heavy-fermion correlations complicate spectroscopic identification, and providing a quantitatively accurate theoretical description for such strongly correlated systems is challenging.

Maldacena proposed that in the large $N$, strong-coupling limit, certain d-dimensional superconformal field theories are exactly dual to classical supergravity (or string theory) on $(d+1)$ dimensional Anti–de Sitter space — the AdS/CFT correspondence \cite{Maldacena:1997re}. This opens a new avenue for studying strongly coupled systems: problems that are intractable in the field theory description can be mapped and solved by comparatively tractable calculations in the dual classical gravitational theory. Because the boundary dynamics are encoded in a higher-dimensional gravitational bulk, the approach is commonly denoted holographic duality.

Several holographic Weyl semimetal models have been proposed \cite{Gursoy:2012ie, Landsteiner:2015lsa, BitaghsirFadafan:2020lkh}. Furthermore, based on these models, the authors have studied anomalous transport phenomena \cite{Furukawa:2024zet, Landsteiner:2015pdh}, topological invariants \cite{Liu:2018djq, Chen:2025akz}, surface states \cite{Ammon:2016mwa}, dislocations \cite{Juricic:2024tbe}, and so on \cite{Juricic:2020sgg, Liu:2020ymx, Ahn:2024ozz, Seo:2025pnl}. A key feature in these models is the anomalous Hall conductivity. Most theoretical treatments assume zero density systems, whereas real materials typically exhibit a finite charge density. Likewise, the characteristic linear dispersion near Weyl points has received relatively little attention in the literature. The linear dispersion of Weyl fermions in the vicinity of the points is a fundamental property of the low-energy quasiparticle excitations in Weyl semimetals. This raises several natural questions: how does the spectral function evolve in a finite density Weyl semimetal? Does a bulk Fermi surface form, and if so, what is its topology? What are the dispersion relations of the low-energy excitations in the finite density model? 

We focus on these questions based on the top-down holographic Weyl semimetal of Ref \cite{BitaghsirFadafan:2020lkh}. This model provides a convenient framework for introducing the chemical potential. It is an extension of the D3/D7 branes constructions \cite{Karch:2002sh}, consisting of an intersection of $N_c$ D3-branes and $N_f$ D7-branes in type IIB supergravity (SUGRA) in ten dimensions. Table \ref{table} summarizes the brane embedding and coordinate assignments. The D3-branes extend along the four spacetime directions $(t, x, y, z)$, while the D7-branes occupy an eight-dimensional worldvolume spanned by $(t, x, y, z, r)$ and wrap an $S^3$. The remaining polar coordinates $(R,\phi)$ parametrize the two-dimensional plane transverse to both the D3 and D7 branes. The dual field theory is a (3+1)-dimensional supersymmetric Yang–Mills theory coupled to probe hypermultiplets; each hypermultiplet contains a Dirac fermion and a pair of complex scalars. The Dirac fermion can be interpreted as a quark or an electron. Building on this construction, Ref \cite{BitaghsirFadafan:2020lkh} implements an axial $U(1)_B$ gauge field $B_j^5$ by choosing $\phi \propto z$ and setting $B_z^5=\frac{\partial_z \phi}{2}$. This results in nonzero Hall conductivities, which are interpreted as the holographic realization of Weyl-semimetal physics.
\begin{table}[h]
  \centering
  {\setlength{\tabcolsep}{10pt}
   \renewcommand{\arraystretch}{1.3}
   \large 
   \resizebox{0.8\textwidth}{!}{%
     \begin{tabular}{l c c c c c c c c c c}
       \toprule
       & $t$ & $x$ & $y$ & $z$ & $r$ &  & $S^3$ &  & $R$ & $\phi$  \\
       \midrule
       D3 & $\times$ & $\times$ & $\times$ & $\times$ & & & & & & \\
       D7 & $\times$ & $\times$ & $\times$ & $\times$ & $\times$ &  $\times$ &  $\times$ &  $\times$ & & \\
       \bottomrule
     \end{tabular}
   }
  }
  \caption{Embedding of branes in the type IIB supergravity background. The D3-branes occupy the worldvolume $ (t, x, y, z)$. The D7-branes coincide with these four spacetime directions and further extend along the radial coordinate $r$ while wrapping an internal $S^3$. The coordinates $R$ and $\phi$ parametrize the directions transverse to both D3-branes and D7-branes.}
  \label{table}
\end{table}

Obtaining the band structure and dispersion relations of Weyl semimetals in our model requires computing fermionic spectral functions, which provide the direct link between theoretical predictions and angle-resolved photoemission spectroscopy (ARPES) measurements. The fermionic spectral function and holographic Fermi surface were initially explored with a probe-fermion technique \cite{Iqbal:2009fd, Liu:2009dm, Lu:2024qxj, Fang:2014jka, Fang:2012pw, Fang:2013ixa}. This method was subsequently applied to holographic Weyl semimetals in Ref \cite{Liu:2018djq} to compute topological invariants via the topological Hamiltonian method. Probe fermions have also been implemented in the D3/D7 framework \cite{Ge:2023ghy, Ammon:2010pg, Kirsch:2006he, Abt:2019tas}. Following these works, we introduce two probe fermions with opposite chirality as in Ref \cite{Liu:2018djq}, and implement them in a bottom-up fashion similar to Ref \cite{Ge:2023ghy}. This choice greatly simplifies the calculations while leaving the essential physical properties unchanged. Our calculated spectral function at zero density indeed exhibits linear dispersion, supporting the concept of a Weyl semimetal. At finite density, we obtain energy bands near the Fermi level and the Fermi surface, observing the Lifshitz transition. This is a phase transition in which the topology of the Fermi surface changes due to a change in the electronic band structure \cite{lifshitz1960anomalies}. It is a crucial concept in condensed matter.

The paper is organized as follows. In Section \ref{sec:Weyl}, we introduce the chemical potential into the holographic flavour brane Weyl semimetal of Ref \cite{BitaghsirFadafan:2020lkh} , thereby constructing a finite density model. In Section \ref{sec:fer}, we add two probe fermions of opposite chirality to this background and compute the corresponding fermionic Green’s functions. In Section \ref{sec:spe}, we present the spectral function and dispersion relation for the zero density system. Section \ref{sec:lif} analyzes the finite density model, examining its dispersion relation and Fermi surface and demonstrating a Lifshitz transition. Finally, Section \ref{sec:dac} summarizes our results and conclusions.

\section{The holographic finite density flavour brane Weyl semimetal}
\label{sec:Weyl}
In this section, we review the holographic flavour brane Weyl semimetal proposed in Ref \cite{BitaghsirFadafan:2020lkh}. Based on this model, we introduce the chemical potential following the method in Ref \cite{Ge:2023ghy} and construct the holographic finite density flavour brane Weyl semimetal. 

The $\mathcal{N}=4$ supersymmetric $SU(N_c)$ Yang-Mills theory (SYM) is holographically dual to type IIB SUGRA in the near-horizon geometry of the $SU(N_c)$ D3-branes, $AdS_5\times S^5$.  To incorporate a nonzero temperature $T$ in the field theory, we replace the $AdS_5$ factor with an $AdS_5$-Schwarzschild black brane, where $T$ is the Hawking temperature of the black brane. The resulting geometry is $AdS_5$-Schwarzschild $\times S^5$. We can write down the metric and four-form $C_4$, which are given by
\begin{align}
    \begin{aligned}
\mathrm{d} s^2 & =\frac{\rho^2}{L^2} \left(-\frac{f(\rho)^2}{h(\rho)} \mathrm{d} t^2+h(\rho) \mathrm{d} \vec{x}^2\right)+\frac{L^2}{\rho^2}\left(\mathrm{~d} r^2+r^2 \mathrm{~d} s_{\mathrm{S}^3}^2+\mathrm{d} R^2+R^2 \mathrm{~d} \phi^2\right), \\
C_4 & =\frac{\rho^4}{L^4} h(\rho)^2 \mathrm{~d} t \wedge \mathrm{~d} x \wedge \mathrm{~d} y \wedge \mathrm{~d} z-\frac{L^4}{\rho^4} r^4 \mathrm{~d} \phi \wedge \omega(\mathrm{~S}^3),
\end{aligned}
\end{align}
where
\begin{align}
    f(\rho)=1-\frac{\rho_h^4}{\rho^4},\quad h(\rho)=1+\frac{\rho_h^4}{\rho^4},\quad \rho^2=r^2+R^2.
\end{align}    
$\rho_h$ is the location of the black brane horizon and $L$ is the $AdS_5$ radius. The black brane's Hawking temperature is given by 
 \begin{align}
     T=\frac{\sqrt{2}}{\pi} \rho_h.
 \end{align}

In the D3/D7 model, the intersection of the $N_f$ D7-branes gives rise to $N_f$ $\mathcal{N}=2$ hypermultiplets introduced in dual boundary SYM \cite{Karch:2002sh}. The D3/D7 model is dual to the (3+1) dimensional SYM theory coupled to hypermultiplets. We also work in the probe limit, $N_f \ll N_c$. The D7-brane action is given by
\begin{align}
    S_{\mathrm{D} 7}=-N_f T_{\mathrm{D} 7} \int d^8 \xi \sqrt{-\operatorname{det}(P[G]+F)}+\frac{1}{2} N_f T_{\mathrm{D} 7} \int P\left[C_4\right] \wedge F \wedge F.
    \label{D7 action}
\end{align}
The first term is the Abelian Dirac–Born–Infeld (DBI) term, and the second is the Wess–Zumino (WZ) term. Note that $T_{D7}$ is the brane tension, $P[G]$ and $P[C_4]$ are the pull back of the metric and four form, and $F=dA$ is the field strength of the $U(1)$  worldvolume gauge field $A$. $\xi$ are the worldvolume coordinates. Different from the original definition of $F$ \cite{polchinskistring}, we have absorbed a factor of $2 \pi \alpha'$ into the $F$.

For constructing a holographic finite density Weyl semimetal, we make the ansatz that 
\begin{align}
    A=A_t,\, R=R(r) \, \text{and} \, \phi=bz.
    \label{ansatz}
\end{align}
Because we work in the probe limit. We can use the form of the induced metric on the D7-branes given in Ref \cite{Furukawa:2024zet}, directly. It is given by
\begin{align}
     ds^{2}_{D7}=L^2\rho^2\left[-\frac{f(\rho)^2}{h(\rho)}\mathrm{d}t^2 + h(\rho)\left(\mathrm{d}x^2 + \mathrm{d}y^2\right) + \left(h(\rho) + \frac{b^2R^2}{\rho^4}\right)\mathrm{d}z^2\right]+\frac{L^2}{\rho^2}[(1+R'^2)\mathrm{d}r^2+r^2\mathrm{d}s^2_{S^3}].
     \label{induced metric}
\end{align}
The field strength $F$ on the D7-brane worldvolume can be expressed as 
\begin{align}
    F_{\alpha \beta}=\partial_{\alpha} A_{\beta}-\partial_{\beta} A_{\alpha},
\end{align}
where $\alpha$ and $\beta$ label the D7 branes worldvolume coordinates $\xi$. In our ansatz \eqref{ansatz}, only the components $F_{rt}=-F_{tr}=\partial_r A_t$ are nonvanishing. As a result, the WZ term vanishes and therefore does not contribute to the D7-brane action. The action \eqref{D7 action} then reduces to the DBI part, which can be written as $S_{D7}=\int \mathcal{L} \mathrm{d}r$. $\mathcal{L}$ is the Lagrangian density of the DBI term. Substituting the induced metric \eqref{induced metric} and $F$'s components into the action \eqref{D7 action}, we can obtain its specific form is given by
\begin{align}
    \mathcal{L}=-\mathcal{N}h(\rho) \, r^3 \sqrt{\left(1+\frac{b^2 L^4 R^2}{h(\rho)\, \rho^4}\right)(-h(\rho) \,A_t'^2+f(\rho)^2\left(1+R'^2\right))}
    \label{Lagrangian density}
\end{align}
where $\mathcal{N}=2 \pi^2 N_f T_{D7}$. Following the stability analysis presented in  Ref \cite{BitaghsirFadafan:2020lkh}, we have set $\phi' = 0$ in \eqref{Lagrangian density}, consistent with the approach adopted therein. If $\phi' \neq 0$, the corresponding conjugate momentum $P_{\phi}$ would be nonzero, potentially resulting in the negative values under the square root in the D7-brane action. This signifies a tachyonic instability in the decay rate \cite{Hashimoto:2013mua, Hashimoto:2014yya}. To circumvent such instability issues, they set $P_{\phi} = 0$ directly in Ref \cite{BitaghsirFadafan:2020lkh}. Hereafter, we set $L=1$ for simplicity. The equation of motion for $A_t$ is $\partial_r (\frac{\mathcal{\partial L}}{\partial A_t'})=0$. So we have
\begin{align}
    \frac{\mathcal{\partial L}}{\partial A_t'}=\mathcal{N}Q .
    \label{NQ}
\end{align}
$Q$ is related to the charge density in the boundary theory. 
We solve \eqref{NQ} and obtain the solution as follows
\begin{align}
     A_t'(r)=-\frac{Q\,f(\rho)\, \rho^2 \sqrt{1+R^{\prime 2}}}{\sqrt{h(\rho)} \sqrt{Q^2 \rho^4+h(\rho)^3 r^6 \rho^4+b^2 h(\rho)^2 L^4 r^6 R^2}}.
     \label{At}
\end{align}
In holography, the boundary theory chemical potential $\mu$ is given by
\begin{align}
    \mu=A_t(\infty)
\end{align}
with boundary condition $A_t(r_h)=0$. $A_t'$ in the Lagrangian density can be eliminated after the Legendre transformation
\begin{align}
    \tilde{\mathcal{L}} \equiv \mathcal{L}-A_t^{\prime}(r) \frac{\partial \mathcal{L}}{\partial A_t^{\prime}(r)}.
    \label{egendre
transformation}
\end{align}
Substituting \eqref{At} into \eqref{egendre
transformation}, the effective Lagrangian density is 
\begin{align}
    \tilde{\mathcal{L}}=-\frac{f(\rho)\sqrt{1+R'^2}\sqrt{\rho^4(Q^2+h(\rho)^3r^6)+b^2 h(\rho)^2r^6R^2}}{\rho^2\sqrt{h(\rho)}}.
    \label{effective Lagrangian}
\end{align}
It will give the equation of motion of $R$ as follows
\begin{align}
    \frac{\partial \tilde{\mathcal{L}}}{\partial R}-\frac{d}{d r}\left(\frac{\partial \tilde{\mathcal{L}}}{\partial R^{\prime}}\right)=0.
\end{align}
When $r \rightarrow \infty$, we have
\begin{align}
    f(\rho) \rightarrow 1 \quad \text{and} \quad h(\rho) \rightarrow1,
\end{align}
where $Q$ in the second root of \eqref{effective Lagrangian} can be ignored. Then the form of he effective Lagrangian density \eqref{effective Lagrangian} becomes the same form in Ref \cite{BitaghsirFadafan:2020lkh} near the boundary. It gives the near-boundary asymptotic expansion of $R(r)$ as follows
\begin{align}
    R(r)=M \left(1-\frac{ b^2}{2 r^2} \log (r)\right)+ \frac{C}{r^2}+\mathcal{O}\left(\frac{\log (r )}{r^4}\right),
\end{align}
with $M$ and $C$ constants. The transverse separation $R$ between the D7- and D3-branes sets the minimal mass of the open strings stretching between them, $\frac{R}{2 \pi \alpha'}$ with $\alpha'$ the string length squared. So $\lim _{r \rightarrow \infty} R(r)=M$ corresponds to a hypermultiplet mass $m=\frac{M}{2 \pi \alpha'}$ in the boundary SYM field theory. 

In our paper, we take the  rescaling as follows
\begin{align}
    R\rightarrow bR, \quad T\rightarrow bT, \quad Q\rightarrow b^3 Q,
\end{align}
and only consider the situation for the low temperature $T=0.05$. The temperature dependence of the holographic spectral function has been extensively investigated, revealing that its peak broadens and diminishes in height with increasing temperature \cite{Ge:2023ghy, Faulkner:2009wj, Faulkner:2011tm, Fang:2015dia}. This is a smearing phenomenon caused by thermodynamic fluctuations.

\begin{figure}[htbp]
    \centering
     \begin{subfigure}{0.48\textwidth}
        \centering
        \includegraphics[width=\textwidth]{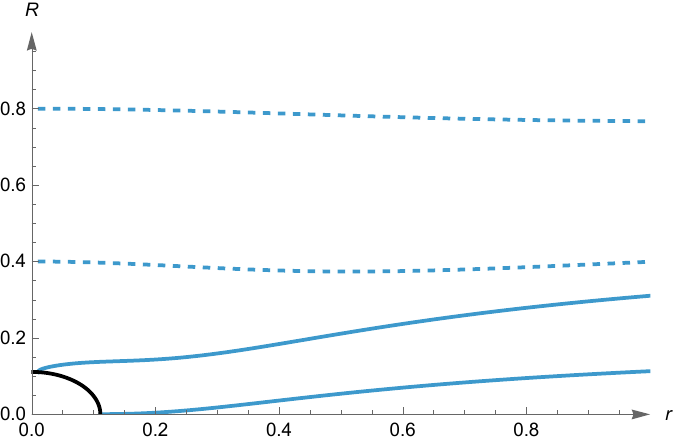}
        \caption{$Q=0$}
        \label{fig:2a}
    \end{subfigure}
    \begin{subfigure}{0.48\textwidth}
        \centering
    \includegraphics[width=\textwidth]{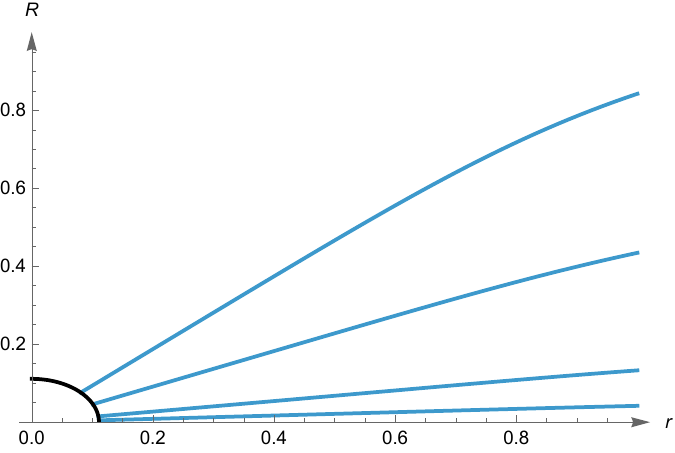}
        \caption{$Q=1$}
        \label{fig:2b}
    \end{subfigure}
    \caption{Embeddings for different IR boundary conditions at $T=0.05$. The black curve indicates the black brane horizon. The blue curves are the embedding functions $R(r)$ for different IR conditions. (a) It corresponds to the zero density case for $Q =0$; the blue dashed curve represents the Minkowski embedding, which describes the insulating phase, while the blue solid curve corresponds to the black hole embedding, characterizing the Weyl semimetal phase. (b) It corresponds to the finite density case for $Q=1$, we find that only the black hole embedding is realized.}
    \label{fig:2}
\end{figure}

Fig.\ref{fig:2} shows the different finite density cases for the embedding function at $T=0.05$. The black curve is the black brane horizon. We set the different regular boundary conditions and plot these blue curves in Fig.\ref{fig:2}. There are two different embedding ways: Minkowski embedding and black hole embedding. The former corresponds to the insulating phase, while the latter corresponds to the Weyl semimetal phase \cite{BitaghsirFadafan:2020lkh}. For black hole embedding, we have the near-horizon condition as follows 
\begin{align}
    R(r)=r \sqrt{\frac{\rho_h^2}{r_h^2}-1} \, \text{and} \, R'(r)=0.
\end{align}
For a given $T$, different values of $r_h$  correspond to different values of $M$. For Minkowski embedding, the embedding function never touch the horizon, and can reach $r=0$ at $R_0>\rho_h$. We have another IR condition as follows 
\begin{align}
    R(r)=R_0 \, \text{and} \, R'(r)=0.
\end{align}
Different values of $R_0$ correspond to different values of $M$. The insulating phase is an inherent feature of the D3/D7 framework, corresponding to a bound state whose spectrum is characterized by a series of discrete Dirac-delta peaks \cite{Mateos:2007vn, Albash:2006bs, Kruczenski:2003be, Karch:2007pd}. In Fig.\ref{fig:2a}, we set $Q=0$ and get a zero density system. There are two classes of Embeddings. In Fig.\ref{fig:2b}, we set $Q=1$ as the finite density case. There is only the black hole embedding way. Because we introduce the chemical potential in this model, it hints that the insulating phase is no longer present. Therefore, this paper focuses exclusively on the Weyl semimetal phase, in both its finite density and zero density cases.

\begin{figure}[htbp]
    \centering    \includegraphics[width=.6\textwidth]{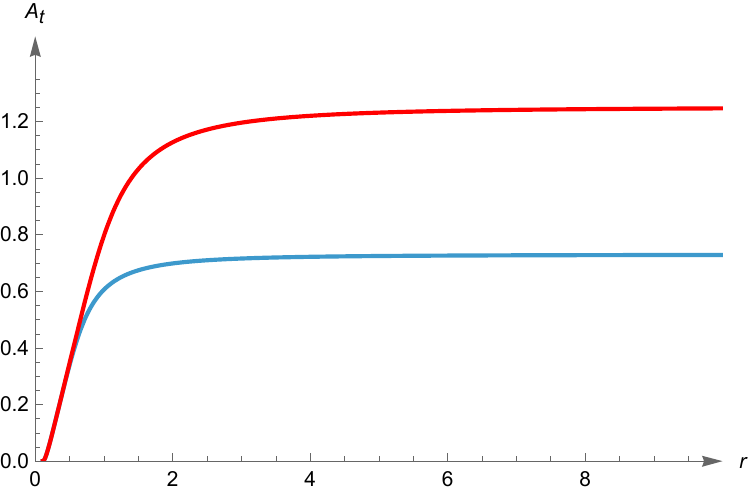}
    \caption{The profile of $A_t$ as a function of $r$ for $T=0.05$ and $M \approx 0.21650$. The blue curve corresponds to $Q=\frac{1}{4}$, while the red curve corresponds to $Q=1$.}
    \label{fig:3}
\end{figure}

Substituting the result for $R(r)$ into the differential equation \eqref{At}, we can obtain the solution function $A_t(r)$. As shown in Fig.\ref{fig:3}, the blue and red curves correspond to $A(t)$ functions for $Q=\frac{1}{4}$ and $Q=1$ with same $M \approx 0.21650$. Both curves converge to a constant at infinity, which corresponds to the boundary chemical potential $\mu$.

At this point, we have constructed a holographic model of a finite density flavored brane system realizing a Weyl semimetal background. At fixed temperatures, the system is fully determined by the embedding function $R$ and the boundary charge density $Q$. The mass parameter $M=\lim _{r \rightarrow \infty} R(r)$, together with $Q$, completely determines the boundary chemical potential $\mu$. Thus, the system can be characterized by the parameters $(M,\mu)$.

\section{Two-probe fermion Setup}
\label{sec:fer}
In this section, we introduce probe fermions on the D7-brane worldvolume to compute the corresponding response function. From this, we obtain the fermionic spectral function and investigate the dispersion relations as well as the Fermi-surface structure of the dual field theory. The probe fermions are introduced in a bottom-up manner, following the approach of Ref \cite{Ge:2023ghy}. Specifically, we consider the fermion action governed by the five-dimensional sector of the induced D7-brane metric \eqref{induced metric},
\begin{align}
     g_{\mu\nu}\mathrm{d}x^{\mu}\mathrm{d}x^{\nu}=\rho^2\left[-\frac{f(\rho)^2}{h(\rho)}\mathrm{d}t^2 + h(\rho)\left(\mathrm{d}x^2 + \mathrm{d}y^2\right) + \left(h(\rho) + \frac{b^2R^2}{\rho^4}\right)\mathrm{d}z^2\right]+\frac{1+R'^2}{\rho^2}\mathrm{d}r^2,
     \label{five induced metric}
\end{align}
and neglect the contribution of the three-sphere, which considerably simplifies the calculation while leaving the fermionic spectral function qualitatively unchanged. In the four-dimensional boundary theory, a bulk Dirac spinor maps to a chiral spinor operator \cite{Iqbal:2009fd}. Accordingly, we introduce two probe spinors, $\Psi_1$ and $\Psi_2$, of opposite chirality to probe the fermionic spectral function relevant for Weyl semimetal physics, following the approach of Refs \cite{Liu:2018djq,Plantz:2018tqf}. The two spinors have masses of opposite sign: taking standard quantization for one and alternative quantization for the other produces boundary operators of opposite chirality.

Two ingredients are crucial for realizing Weyl physics. The first, $B$, breaks time-reversal symmetry but preserves axial symmetry. The second, $\Phi$, explicitly breaks axial symmetry and enables spinors of opposite chirality to couple to one another. These terms lead to the following action for the probe spinors:
\begin{align}
    \begin{aligned}
S & =S_1+S_2+S_{\mathrm{int}}, \\
S_1 & =\int d^5 x \sqrt{-g} i \bar{\Psi}_1\left(\Gamma^a D_a-m_f-i qB_a \Gamma^a\right) \Psi_1, \\
S_2 & =\int d^5 x \sqrt{-g} i \bar{\Psi}_2\left(\Gamma^a D_a+m_f+i qB_a \Gamma^a\right) \Psi_2, \\
S_{\mathrm{int}} & =-\int d^5 x \sqrt{-g}\left(i \eta \Phi \bar{\Psi}_1 \Psi_2+i \eta^* \Phi^* \bar{\Psi}_2 \Psi_1\right),
\end{aligned}
\label{probe}
\end{align}
where $ D_a=\partial_a+\frac{1}{8} \omega_{\underline{m n}, a} [\Gamma^{\underline{m}},\Gamma^{\underline{n}}]-iqA_a$, $g$ is the determinant of the five-dimensional part of the induced metric and $\eta$ is the coupling constant. We just set the coupling constant $\eta=1$. Note that $\Gamma^{\underline{a}}$ is the Gamma matrix in the tangent space and $\Gamma^a$ is the Gamma matrix in the curved spacetime. As the metric \eqref{five induced metric} is diagonal, a special choice of vielbein is possible. Therefore the curved spacetime Gamma matrices satisfy the following relations with the tangent space Gamma matrices:
\begin{align}
\Gamma^t=\frac{\Gamma^{\underline{t}}}{\sqrt{-g_{tt}}}, \, \Gamma^r=\frac{\Gamma^{\underline{r}}}{\sqrt{g_{rr}}}, \, \Gamma^x=\frac{\Gamma^{\underline{x}}}{\sqrt{g_{xx}}}, \, \Gamma^z=\frac{\Gamma^{\underline{z}}}{\sqrt{g_{zz}}}.
\label{elation }
\end{align}

In our holographic finite density flavour brane Weyl semimetal model, $B_z=\frac{\partial_z \phi}{2}$ and $\Phi=R$. We can obtain the equations of motion
\begin{align}
    \begin{aligned}
& \left(\Gamma^a D_a-m_f-i qB_z \Gamma^z\right) \Psi_1- R \Psi_2=0, \\
& \left(\Gamma^a D_a+m_f+i qB_z \Gamma^z\right) \Psi_2- R  \Psi_1=0.
\end{aligned}
\label{motion}
\end{align}
We use the following transformation,
 \begin{align}
     \Psi_l=(-gg^{rr})^{-1 / 4} \psi_l e^{-i \omega t+i k_x x+i k_y y+i k_z z}, \quad l=1,2,
     \label{transform}
 \end{align}
to eliminate the spin connection and transform into momentum space. Substituting \eqref{transform} and \eqref{elation } into \eqref{motion}, the equations of motion become

\begin{align}
\left(
    \sqrt{g^{rr}}\Gamma^{\underline{r}}\partial_r 
    +  \left( -i(\omega+q A_t) \sqrt{-g^{tt}}\Gamma^{\underline{t}} + i \sqrt{g^{xx}}k_x \Gamma^{\underline{x}} \right)
    + \sqrt{g^{zz}} \left( i(k_z \mp qB_z) \Gamma^{\underline{z}}\right)
   \mp m_f
\right) \psi_{l}
-  R \psi_{\bar{l}}
= 0,
\label{Dirac equation}
\end{align}
where $B_z=\frac{b}{2}$. We adopt the same $\Gamma$‑matrix representation as those in \cite{Liu:2018djq}, specified here for completeness
\begin{align}
    \Gamma^{\underline{t}}=\left(\begin{array}{ll}
0 & i \\
i & 0
\end{array}\right), \quad \Gamma^{\underline{i}}=\left(\begin{array}{cc}
0 & i \sigma^i \\
-i \sigma^i & 0
\end{array}\right), \quad \Gamma^{\underline{r}}=\left(\begin{array}{cc}
1 & 0 \\
0 & -1
\end{array}\right) .
\end{align}

Near the boundary $r\rightarrow \infty$, the Dirac fields behaves as
\begin{align}
    \psi_1=\left(\begin{array}{cc}
a_1^1 & r^{m_f}+\cdots \\
a_2^1 & r^{m_f}+\cdots \\
a_3^1 & r^{-m_f}+\cdots \\
a_4^1 & r^{-m_f}+\cdots
\end{array}\right), \quad \psi_2=\left(\begin{array}{cc}
a_1^2 & r^{-m_f}+\cdots \\
a_2^2 & r^{-m_f}+\cdots \\
a_3^2 & r^{m_f}+\cdots \\
a_4^2 & r^{m_f}+\cdots
\end{array}\right) .
\end{align}
Our goal is to compute the boundary retarded Green’s function. Because $\psi_1$ and $\psi_2$ obey opposite quantization conditions, they carry opposite chirality and are coupled to each other. Consequently, the left-hand component of $\psi_1$ and the right-hand component of $\psi_2$ serve as the sources, whereas the right-hand component of $\psi_1$ and the left-hand component of $\psi_2$ correspond to the expectation values. We can therefore arrange the sources and expectation values into two matrices as follows
\begin{align}
    S=\left(\begin{array}{cccc}
a_1^{1, I} & a_1^{1, I I} & a_1^{1, I I I} & a_1^{1, I V} \\
a_2^{1, I} & a_2^{1, I I} & a_2^{1, I I I} & a_2^{1, I V} \\
a_3^{2, I} & a_3^{2, I I} & a_3^{2, I I I} & a_3^{2, I V} \\
a_4^{2, I} & a_4^{2, I I} & a_4^{2, I I I} & a_4^{2, I V}
\end{array}\right) \text { and } E=\left(\begin{array}{cccc}
-a_1^{2, I} & -a_1^{2, I I} & -a_1^{2, I I I} & -a_1^{2, I V} \\
-a_2^{2, I} & -a_2^{2, I I} & -a_2^{2, I I I} & -a_2^{2, I V} \\
a_3^{1, I} & a_3^{1, I I} & a_3^{1, I I I} & a_3^{1, I V} \\
a_4^{1, I} & a_4^{1, I I} & a_4^{1, I I I} & a_4^{1, I V}
\end{array}\right) .
\end{align}
The retarded Green’s function is obtained from 
\begin{align}
   G=i \Gamma^{\underline{t}} ES^{-1}. 
   \label{G}
\end{align}
By rearranging and combining \eqref{Dirac equation}, the Dirac equation can be cast as a relation between the source matrix $S$ and the expectation matrix $E$. The explicit form is
\begin{align}
    \partial_r S =-A_{+}E+m \sqrt{g_{rr}} S, \label{S}\\
    \partial_r E =A_{-}S-m \sqrt{g_{rr}} E,\label{E}
\end{align}
where 
\begin{align}
    A_{\pm}=\left(
\begin{array}{cc}
\pm R\sqrt{g_{rr}} & v_--w_{\mp} \\[6pt]
v_++w_{\pm} & \pm R\sqrt{g_{rr}}
\end{array}
\right),
v_{\pm}=(\omega+A_t) \frac{\sqrt{g_{rr}}}{\sqrt{-g_{tt}}} \pm k_x \frac{\sqrt{g_{rr}}}{\sqrt{g_{xx}}} \sigma_1,
w_{\pm}= (k_z \pm B_z) \frac{\sqrt{g_{rr}}}{\sqrt{g_{zz}}} \sigma_3.
\end{align}

Taking the derivative of both sides of the definition of the Green’s function \eqref{G} yields the following result:
\begin{align}
    \partial G &= i \Gamma^{\underline{t}} (\partial E)S^{-1}+i \Gamma^{\underline{t}} E \partial (S^{-1}) \\
    &=i \Gamma^{\underline{t}} (\partial E)S^{-1}-i \Gamma^{\underline{t}}ES^{-1}(\partial S)S^{-1}. \label{eq}
\end{align}
Substituting \eqref{S}, \eqref{E}, and \eqref{G} into equation \eqref{eq} and simplifying yields the flow equation for the Green's function $G$, which can be written as
\begin{align}
    \partial G= i \Gamma^{\underline{t}} A_--2 m \sqrt{g_{rr}} G + i  G A_+  \Gamma^{\underline{t}} G.
    \label{flow}
\end{align}
At low temperature, as $r \rightarrow r_h$ the off-diagonal components of $A_{\pm}$ dominate over the diagonal ones. For the flow equation \eqref{flow}, the near-horizon boundary condition is $G(r_h)=i \times 1_{4\times 4}$ as in Refs \cite{Plantz:2018tqf,Yuk:2022lof}. Integrating this flow equation from the horizon to the boundary then yields the boundary retarded Green’s function. In the next section, we compute the spectral function, which is defined as $A(k,\omega)=\operatorname{Im}[\operatorname{tr} G (k,\omega)]$.

\section{Spectral function}
\label{sec:spe}
We first examine the zero density case and then proceed to the finite density case. The fermionic spectral function is obtained numerically from the flow equation \eqref{flow}. Based on these results, we analyze the dispersion relation and the Fermi surface structure of the holographic finite density flavour brane Weyl semimetal. Throughout the analysis, we fix the temperature at $T=0.05$, set the probe fermion mass to $m_f=0$, and choose the probe fermion charge as $q=1$. The effects of these parameters have been extensively studied: higher temperature and smaller $m_f$ tend to broaden the spectral function peak, while larger $q$ can give rise to multiple Fermi surfaces \cite{Fang:2015dia, Faulkner:2009wj}. In this work, our main interest lies in the dynamics of the boundary field theory itself. Accordingly, we focus on the influence of the chemical potential $\mu$ and the mass parameter $M$ of the coupled fermions in the boundary field theory on the dispersion relation and Fermi surface of this system. Here, we adjust both $Q$ and $M$ to control the chemical potential.

Starting with the zero density case, $Q=0$, the chemical potential vanishes, $\mu=0$, and the system exhibits particle–hole symmetry. Under these conditions, the model reduces to the original holographic flavour brane description of a Weyl semimetal \cite{BitaghsirFadafan:2020lkh}. This system exhibits a phase transition between the Weyl semimetal phase and an insulating phase, controlled by the parameter $R$. Since the system is at zero density, no Fermi surface is present. In this work, we focus exclusively on the Weyl semimetal phase.

\begin{figure}[htbp]
    \centering    \includegraphics[width=.6\textwidth]{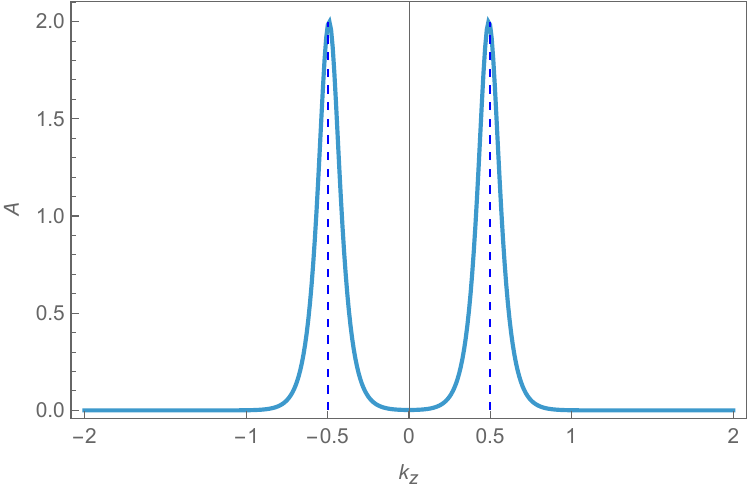}
    \caption{The spectral function $A(k_z)$ for $T=0.05$, $M \approx 0.02$, and $\omega=0$. The spectral function $A$ exhibits a finite weight contribution at $k_z=\pm \frac{1}{2}$.}
    \label{fig:}
\end{figure}

In the study of Ref \cite{BitaghsirFadafan:2020lkh}, the system was identified as a Weyl semimetal when $M < M_{\text{critical}}$. Accordingly, we begin with a relatively small value of $M \approx 0.02$. Since our model incorporates an axial coupling $B_z^5$ along the $z$-direction, we first compute and analyze the spectral function at $\omega = 0$ as a function of $k_z$. As shown in Fig.\ref{fig:}, the spectral function exhibits nonzero spectral weight near $k_z = \pm b = \pm \frac{1}{2}$, indicating the presence of Weyl points.

\begin{figure}[htbp]
    \centering    \includegraphics[width=.6\textwidth]{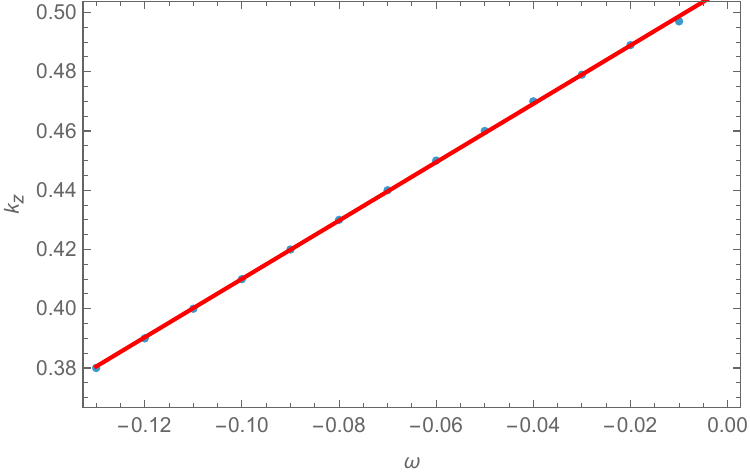}
    \caption{The relation $k_z-\omega$ near a Weyl point. The red line has been fitted to the blue data points. It gives a linear dispersion relation. }
    \label{fi}
\end{figure}

Furthermore, we analyze the dispersion relation of quasiparticle excitations in the vicinity of the Weyl point. As shown in Fig.\ref{fi}, the nontrivial spectral-weight contributions of these excitations are extracted from the spectral function by sampling $\omega$ in steps of $0.01$ over the range $\omega\in(-0.13, 0)$. The red solid line represents the dispersion relation obtained by fitting the resulting data points. The fitted dispersion is given by
\begin{align}
    \omega \approx 1.02 k_z -0.52.
\end{align}
We find that the low energy excitations near the Weyl point exhibit a linear dispersion. The fitted dispersion indicates that at $\omega$ the momentum is $ k_z \approx \frac{1}{2}$, which corresponds to the Weyl point location.

\begin{figure}[htbp]
    \centering    \includegraphics[width=.6\textwidth]{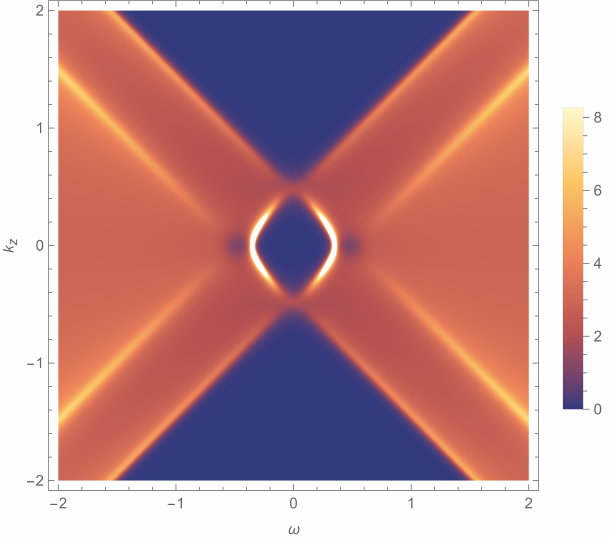}
    \caption{The density plot of the spectral function $A(\omega, k_z)$ for $T=0.05$ and $M \approx 0.02$. The figure reveals four energy bands with a linear dispersion near the Weyl points.}
    \label{fig:4}
\end{figure}

To reveal the global energy band structure, we compute the fermionic spectral function as a function of $k_z$ and $\omega$, the result is shown in Fig.\ref{fig:4}. Two such points are located at $(k_z, \omega) = (\pm \frac{1}{2}, 0)$, consistent with the positions identified in Fig.\ref{fig:}. The figure clearly reveals the linear Weyl cone near the Weyl points, along with the presence of four bands. The band structure shown in Fig.\ref{fig:4} matches that of a conventional Weyl semimetal as described by the field theory illustrated in Fig.\ref{fig:1a}. For sufficiently small $M$, the system can be treated as a specific type of free Dirac fermion system, analogous to the theoretical model given in \eqref{Lagrangian}, which accounts for the linear dispersion. This behavior has also been experimentally observed in TaAs materials \cite{lv2015experimental,xu2015discovery}. Notably, the holographic flavour brane Weyl semimetal exhibits a finite spectral weight at the Weyl point. Since the axial anomaly coupling $B_z^5$ is a constant $b$, it has been appropriately rescaled. In this scenario, the Hall conductivity, as discussed in \cite{BitaghsirFadafan:2020lkh}, shows no dependence on $m$. Nevertheless, stable quasiparticle excitations persist in the vicinity of the Weyl point.

\begin{figure}[htbp]
    \centering
    \begin{subfigure}{0.45\textwidth}
        \centering
        \includegraphics[width=\textwidth]{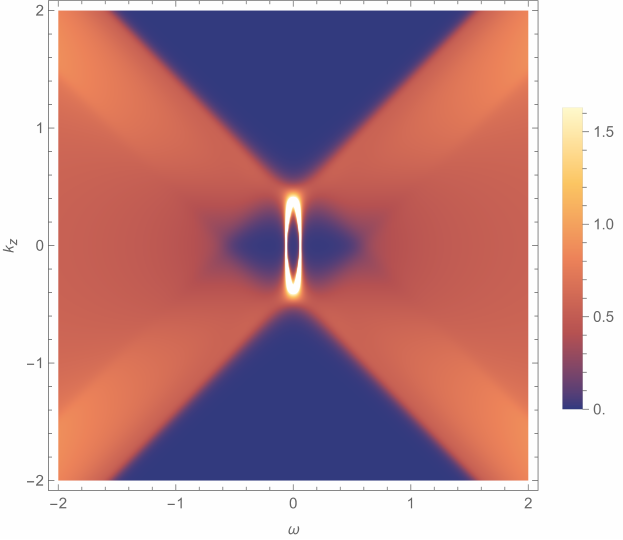}
        \caption{$M \approx 0.06$}
        \label{fig:5a}
    \end{subfigure}
    \begin{subfigure}{0.48\textwidth}
        \centering
    \includegraphics[width=\textwidth]{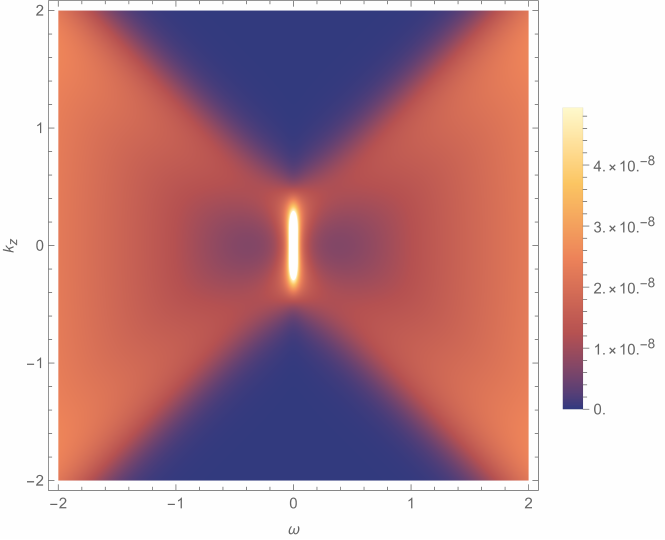}
        \caption{$M \approx M_{\text{critical}}$}
        \label{fig:5b}
    \end{subfigure}
    \caption{(a) The density plot of the spectral function $A(\omega, k_z)$ for $M \approx 0.06$ and $T=0.05$. (b) The density plot of the spectral function $A(\omega, k_z)$ for $M \approx M_{\text{critical}}$ and $T=0.05$. As $M$ increases to $M_{\text{critical}}$, the spectral weight decreases rapidly, and the dispersion between the Weyl points becomes compressed.}
    \label{fig:5}
\end{figure}

When $M$ is increased slightly to $M \approx 0.06$, as shown in Fig.\ref{fig:5a}, the spectral function becomes broadened across all four bands. No clear excitations remain outside the Weyl points, while a sharp peak persists between them. The dispersion relation connecting the two Weyl points becomes compressed and deviates from linearity. As depicted in Fig.\ref{fig:5b}, the spectral function is almost entirely suppressed as $M$ approaches $M_{\text{critical}}$. Once $M$ exceeds $M_{\text{critical}}$, the system undergoes a quantum phase transition from the Weyl semimetal phase to an insulating phase, consistent with the description in \cite{BitaghsirFadafan:2020lkh}. Although the dispersion between the Weyl points is visibly compressed, the scale bar in the figure indicates that the spectral weight is significantly diminished. While some residual Weyl characteristics may still be discernible, the system more closely resembles a gapped band structure.

In the boundary field theory, the fermion mass is given by $m=\frac{M}{\sqrt{\lambda}}$. When the mass parameter $m$ is sufficiently small, the model recovers the characteristic band structure of a conventional Weyl semimetal, as presented in Fig. \ref{fig:4}. This structure has received direct experimental confirmation \cite{lv2015experimental,xu2015discovery}. As the mass $m$ increases, the peak in the spectral function decreases rapidly. Nevertheless, stable quasiparticle excitations persist between the Weyl points; their dispersion relation becomes increasingly compressed. Approaching the critical mass, the bands are nearly squeezed together, and the spectral features outside the Weyl points become broadened and less distinct. This influence of 
$m$ on the spectral function and band structure stems from the strongly coupled nature of the holographic Weyl semimetal model.

\section{Lifshitz transition}
\label{sec:lif}
We next analyze the spectral function at finite charge density. By inspecting the band structure and the topology of the Fermi surface, we find that the system undergoes a Lifshitz transition when the chemical potential $\mu$ or the mass parameter $M$ is varied. The notion of a Lifshitz transition was introduced by I. M. Lifshitz in 1960 to describe an abrupt change in the topology of the Fermi surface that occurs without symmetry breaking \cite{lifshitz1960anomalies}. Consequently, it cannot be characterized by a local order parameter. Such transitions may be driven by external control parameters, temperature, magnetic field, pressure, doping, interactions, etc. As control parameters vary continuously, a local saddle point of the band structure crosses the Fermi level, inducing topological changes of the Fermi surface in momentum space, such as the appearance or disappearance of pockets and transitions between connected and disconnected Fermi surfaces.

We concentrate first on the influence of the chemical potential. So we keep $M \approx 7.8 \times 10^{-3}$ and set $Q=\frac{1}{4}$ and $\frac{1}{2}$. The former corresponds to the system with $\mu \approx 0.73$. The latter corresponds to the system with $\mu \approx 0.96$. They are both with a small fermion mass $m=\frac{M}{\sqrt{\lambda}}$. The key distinction from the zero density case is that a finite density system leads to the formation of a Fermi surface.

\begin{figure}[htbp]
    \centering
    \begin{subfigure}{0.48\textwidth}
        \centering        \includegraphics[width=\textwidth]{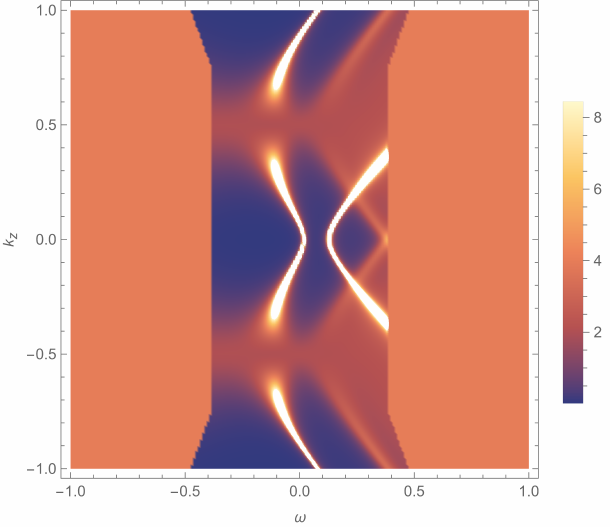}
        \caption{$Q=\frac{1}{4}$}
        \label{fig:6a}
    \end{subfigure}
    \begin{subfigure}{0.48\textwidth}
        \centering
    \includegraphics[width=\textwidth]{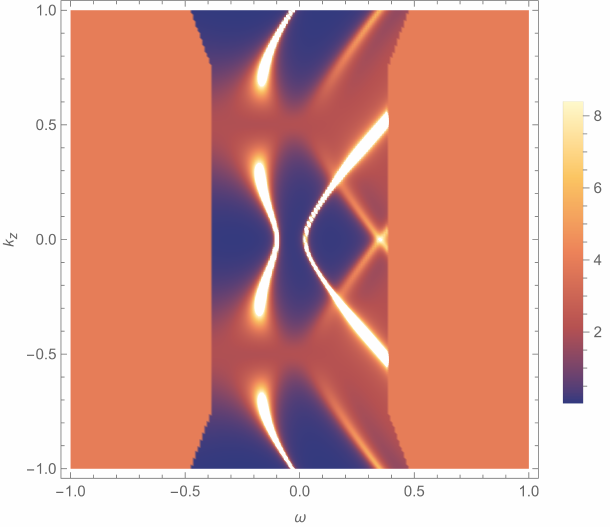}
        \caption{$Q=\frac{1}{2}$}
        \label{fig:6b}
    \end{subfigure}
    \caption{The density plot of the spectral function $A(\omega,k_z)$ for the holographic finite density Weyl semimetal at $ M\approx 7.8\times 10^{-3}$ and $T=0.05$. (a) For $Q=\frac{1}{4}$, it corresponds to the chemical potential $\mu \approx 0.73$. (b) For $Q=\frac{1}{2}$, it corresponds to the chemical potential $\mu \approx 0.96$. The figure displays two energy bands, with the left band passing through two Weyl points and intersecting $\omega=0$ to define the Fermi momentum. As the chemical potential increases from $\mu\approx0.73$ to $\mu\approx0.96$, the position of $\omega=0$ shifts, manifesting as a leftward displacement of the energy band. }
    \label{fig:6}
\end{figure}

We obtain the band structure for the holographic finite density flavour brane Weyl semimetal, as shown in Fig.\ref{fig:6}. Since holographic calculations determine the bands only in the vicinity of $\omega = 0$, much like the spectral function obtained from ARPES experiments, they generally do not provide the complete band structure. Only two bands are visible in Fig.\ref{fig:6}. Our spectral function results are consistent with experimental ARPES data, which typically show signal below the Fermi energy \cite{lv2015experimental,xu2015discovery}. The left band crosses two Weyl points. The finite spectral weight at the Weyl point is consistent with the findings in the previous section. The Weyl points remain located at $k_z = \frac{1}{2}$, though not at $\omega = 0$. Unlike the zero density situation, the intersection of the left band with $\omega = 0$ determines the Fermi momentum. As we increase $\mu$, the entire band structure shifts downward in $\omega$. It can be seen from Fig.\ref{fig:6} that $\mu$ cannot be arbitrarily large: too high a value of $\mu$ would cause the right band to cross $\omega = 0$, which is inconsistent with fundamental concepts in condensed matter physics.

\begin{figure}[htbp]
    \centering
    \begin{subfigure}{0.48\textwidth}
        \centering        \includegraphics[width=\textwidth]{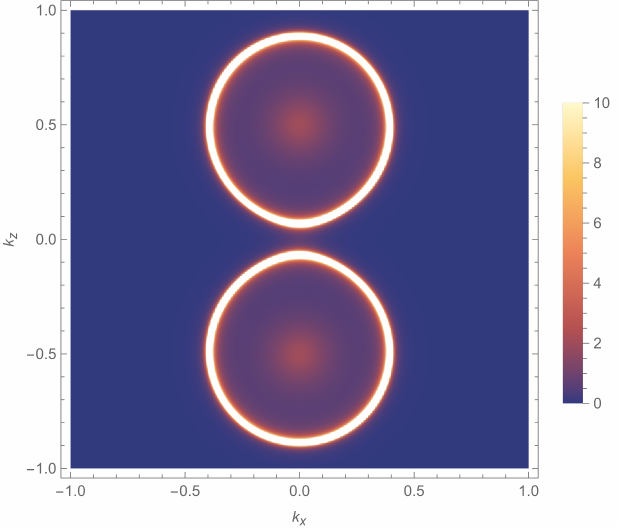}
        \caption{$Q=\frac{1}{4}$}
        \label{fig:7a}
    \end{subfigure}
    \begin{subfigure}{0.48\textwidth}
        \centering
    \includegraphics[width=\textwidth]{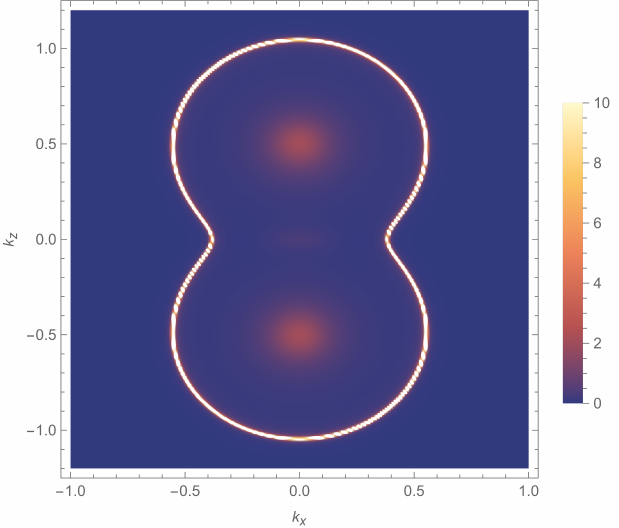}
        \caption{$Q=\frac{1}{2}$}
        \label{fig:7b}
    \end{subfigure}
    \caption{The density plot of the spectral function $A(k_x,k_z)$ for the holographic finite density Weyl semimetal at $ M\approx 7.8\times 10^{-3}$, $T=0.05$ and $\omega=0$. (a) For $Q=\frac{1}{4}$, it corresponds to the chemical potential $\mu \approx 0.73$. (b) For $Q=\frac{1}{2}$, it corresponds to the chemical potential $\mu \approx 0.96$. The figure clearly reveals the Fermi surface in the $(k_x,k_z)$ momentum space. At $\mu \approx 0.73$, two small Fermi pockets appear, each enclosing a different Weyl point. As the chemical potential increases to $\mu \approx 0.96$, these pockets merge into a single large Fermi surface that simultaneously encloses both Weyl points. This behavior is a clear signature of a Lifshitz transition.}
    \label{fig:7}
\end{figure}

In Fig.\ref{fig:6}, the Fermi momentum can be identified, indicating the presence of a Fermi surface in the system. To examine the structure of the Fermi surface more clearly and intuitively, we fix $k_y = 0$ and set $\omega = 0$, thereby studying the momentum space cut in the $k_x$–$k_z$ plane. As shown in Fig.\ref{fig:7}, the white outline corresponds to a sharp spectral peak, representing the Fermi surface. The spectral function exhibits nonzero intensity near $k_z = \pm \frac{1}{2}$ at $k_y = 0$, indicating the locations of the Weyl points. For $\mu \approx 0.73$, two distinct Fermi pockets emerge, each enclosing one Weyl point as shown in Fig.\ref{fig:7a}. As $\mu$ increases, these two Fermi pockets merge into a single large Fermi surface that surrounds both Weyl points, as depicted in Fig.\ref{fig:7b}. This evolution signifies a topological Lifshitz transition. Quantitatively, we can see in Fig.\ref{fig:6} that the increase in chemical potential the left band's intersections with $\omega=0$ undergoes an abrupt, discontinuous transition from four to two. It is this change in the electronic band structure that leads to the Lifshitz transition. In topological semimetals, such a transition alters the topological invariants associated with the Fermi surface, which in turn can lead to many different types of Lifshitz transition  \cite{Volovik:2016dsj}. Hence, this is referred to as a topological Lifshitz transition.

\begin{figure}[htbp]
    \centering
    \begin{subfigure}{0.48\textwidth}
        \centering        \includegraphics[width=\textwidth]{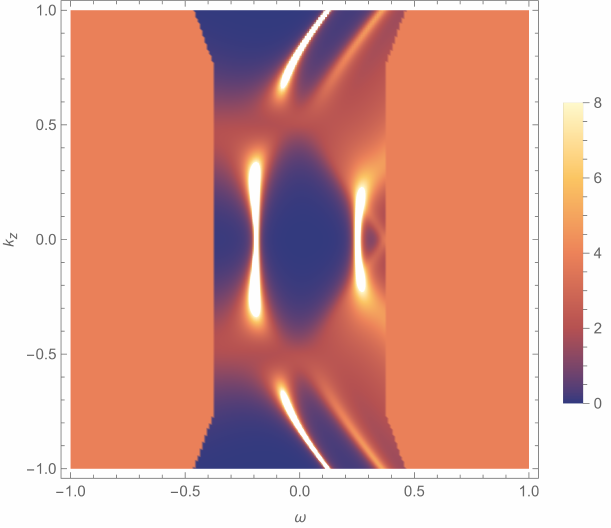}
        \caption{}
        \label{fig:8a}
    \end{subfigure}
    \begin{subfigure}{0.48\textwidth}
        \centering
    \includegraphics[width=\textwidth]{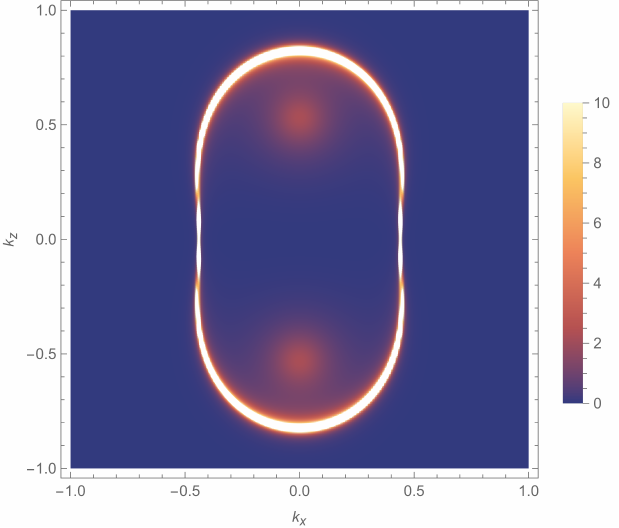}
        \caption{}
        \label{fig:8b}
    \end{subfigure}
    \caption{(a) The density plot of the spectral function $A(\omega,k_z)$ for the holographic finite density Weyl semimetal at $ M\approx 7.1\times 10^{-2}$ and $Q=\frac{1}{4}$. (b) The density plot of the spectral function $A(k_x,k_z)$ for the holographic finite density Weyl semimetal at $ M\approx 7.1\times 10^{-2}$, $Q=\frac{1}{4}$ and $\omega=0$. Keeping $T=0.05$ fixed, the boundary chemical potential for the parameters above is $\mu\approx0.73$. Compared with Fig.\ref{fig:6a} and Fig.\ref{fig:7a}, we increase $M$ to $7.1 \times 10^{-2}$. On the one hand, we observe a drastic change in the band shape. On the other hand, in momentum space, this manifests as the merging of two Fermi pockets into a single large Fermi surface. This also signifies a Lifshitz transition, though it is distinct from the one driven by $\mu$.}
    \label{fig:8}
\end{figure}

We now investigate the influence of the parameter $M$ on the spectral function and the Fermi surface. Setting $M \approx 7.1 \times 10^{-2}$ and $Q = \frac{1}{4}$, the chemical potential of the system remains approximately unchanged at $0.73$. The resulting spectral function and Fermi surface are displayed in Fig.\ref{fig:8}.

As shown in Fig.\ref{fig:8a}, at $ k_z=\pm \frac{1}{2}$, a finite spectral weight is observed, indicating the presence of Weyl points. However, in comparison with Fig.\ref{fig:6a}, the band between the two Weyl points becomes notably flatter. Meanwhile, Fig.\ref{fig:8b} reveals a large, connected Fermi surface, which contrasts with the two separate Fermi pockets illustrated in Fig.\ref{fig:7a}. This suggests that a slight increase in $M$ induces a topological change in the Fermi surface—namely, the merger of two Fermi pockets into a single large Fermi surface.

This $M$-driven Lifshitz transition is fundamentally distinct from the $\mu$-driven  transition shown in Fig.\ref{fig:7}. In the latter case, as illustrated in Fig.\ref{fig:6}, an increase in $\mu$ shifts the two energy bands to the left, reducing the number of intersection points with $\omega=0$ from four to two, thereby triggering the transition. In contrast, a comparison between Fig.\ref{fig:6a} and Fig.\ref{fig:8a} demonstrates that increasing $M$ flattens the band between the Weyl points, which similarly reduces the number of $\omega=0$ intersections from four to two and results in a Lifshitz transition, albeit through a different mechanism.

We observe the Lifshitz transition via probe fermions in the holographic finite density flavour brane Weyl semimetal model. This model features two free parameters: a chemical potential $\mu$ and a mass parameter $M$, the latter corresponding to the mass of the coupled fermions in the boundary field theory. Increasing the chemical potential shifts the Fermi level. In Fig.\ref{fig:6}, the Fermi level is at $\omega=0$; its shift is reflected in the corresponding displacement of the energy bands. Concurrently, the topology of the Fermi surface changes from two separate pockets to a single connected one. Altering the chemical potential is one of the most direct and prevalent methods to drive a Lifshitz transition, as it displaces the Fermi level and modifies the Fermi surface.

On the other hand, varying the mass parameter $M$ also induces a Lifshitz transition, but through a different mechanism. Unlike the chemical potential, changes in $M$ do not shift the Fermi level. Instead, they modify the band shape—specifically, causing the band connecting the two Weyl points to become increasingly flattened. This effect is already evident at zero density situation. Such flattening alters the band structure near the Weyl points, thereby changing the geometry of the Fermi surface and triggering a Lifshitz transition.

The Lifshitz transition has been experimentally observed in a variety of materials, driven by multiple factors including temperature, magnetic field, pressure, and interaction. In certain iron-based superconductors, this transition is closely linked to superconductivity and the critical temperature \cite{liu2010evidence,ren2017superconductivity}. Weyl semimetals also exhibit a Lifshitz transition under the influence of Coulomb interactions \cite{xu2018evidence}. Such correlation-induced effects are analogous to the $M$-driven Lifshitz transition in our model, which does not correspond to a rigid band shift but rather reflects strong coupling behavior. Conversely, nearly rigid band shifts have been achieved in Weyl semimetals through specific doping techniques, leading to the detection of electron pocket emergence or disappearance \cite{lohani2023electronic}—a phenomenon similar to our $\mu$-driven Lifshitz transition. Notably, different Lifshitz transitions can be induced in the same Weyl semimetal by tuning various parameters \cite{jung2024quantum}. Similarly, our model supports both $M$-driven and $\mu$-driven Lifshitz transitions.

Only the electronic structure in the immediate vicinity of the Fermi surface governs a material’s transport and thermodynamic behaviour. Lifshitz transitions, changes in the topology of the Fermi surface, therefore have profound effects on these properties. They are intimately linked to the formation of charge and spin density waves \cite{lin2010lifshitz}, can influence superconducting pairing through van Hove singularities \cite{chen2012lifshitz}, and may drive Pomeranchuk-type Fermi-surface distortions \cite{carr2010lifshitz}. In Weyl semimetals, Lifshitz transitions modify the topological invariants that characterize Fermi pockets and so can strongly affect electronic transport; for example, they have been associated with negative magnetoresistance and related phenomena \cite{yang2019topological}. Because Lifshitz transitions also can appear in strongly correlated systems \cite{bragancca2018correlation}, holographic approaches are a natural framework for their study. Our holographic finite density flavour brane model describes a strongly coupled Weyl semimetal and therefore provides a particularly suitable toy model in which Lifshitz transitions can be realized and investigated directly.

The Lifshitz transition exerts a profound influence on Weyl semimetals, significantly affecting the evolution of Fermi arcs and surface states. It further modifies charge transport properties and is often accompanied by topological phase transitions, which involve the annihilation or creation of Weyl points and lead to changes in topological invariants. This transition also alters key transport coefficients, such as the Hall effect, thermoelectric response, and magnetoresistance. By reconstructing the Fermi surface, the Lifshitz transition impacts essential material properties that are largely governed by low-energy electronic behavior near the Fermi level. As such, it represents a critical phenomenon in condensed matter physics. 

\section{Discussion and Conclusion}
\label{sec:dac}
We studied the holographic flavour brane Weyl semimetal of Ref \cite{BitaghsirFadafan:2020lkh} by introducing two probe fermions of opposite chirality and computing the retarded Green’s function and spectral function. At small mass parameter $M$, the spectral function exhibits the hallmark signatures of a Weyl semimetal, a pair of Weyl points and linear dispersion, thereby providing spectral (ARPES-like) evidence that complements the transport-based conclusions of Ref \cite{BitaghsirFadafan:2020lkh}. Extending the model to finite chemical potential $\mu$, we find the emergence of Fermi surfaces and a Lifshitz transition: two disconnected pockets around the Weyl points merge into a single large Fermi surface. Importantly, the Lifshitz transition can occur via two distinct mechanisms, a rigid energy shift induced by $\mu$ or a compression of the inter-point dispersion driven by $M$, reflecting different origins of Fermi surface reconstruction in the holographic dual. 

There are several directions in which this work can be further developed. In holographic realizations of Weyl semimetals, one important aspect that remains to be addressed is the topological structure and the associated surface Fermi arcs, together with their chirality and transport signatures, for example, negative magnetoresistance. Within the present model, we do not observe gap opening, a limitation that likely stems from the model’s built-in assumptions and therefore requires additional physical ingredients to capture gap-forming mechanisms.  In our holographic spectral analysis, we employ two probe fermions and thereby obtain two energy bands. How should probe fermions be implemented in the holographic model to correspond to the different ARPES measurements used in the experiment? Matching experimental and theoretical results remains a central challenge. And deriving transport coefficients from holographic spectral functions is a crucial and nontrivial step that demands theoretical self-consistency.

In summary, these results demonstrate a robust, strongly coupled realization of Weyl physics in the flavour brane construction and highlight spectral probes as a direct diagnostic of holographic topological phases. We observe the Lifshitz transition in the holographic model. This is an important concept in condensed matter physics proposed in the 60s of the last century, and there are already a large number of experiments and theories closely related to it. Gauge/gravity duality is indeed a natural and perfect platform for studying strongly coupled systems.

\acknowledgments

We would like to thank Xiao Hu and Shuta Ishigaki for helpful discussions. The study was partially supported by NSFC, China (grant No.12275166 and  No.12311540141). This work is also supported by NRF of Korea with grant No.NRF-2021R1A2B5B02002603, RS-2023-00218998 and NRF-2022H1D3A3A01077468.



\begin{thebibliography}{10}

\bibitem{yan2017topological}
B.~H. Yan and C.~Felser.
\newblock Topological materials: Weyl semimetals.
\newblock {\em Annu. Rev. Condens. Matter Phys}, 8(1):337--354, 2017.

\bibitem{weyl1929electron}
H.~Weyl.
\newblock Electron and gravitation.
\newblock {\em z. Phys}, 56:330--352, 1929.

\bibitem{lv2015experimental}
B.~Q. Lv, H.~M. Weng, B.~B. Fu, X.~P. Wang, H.~Miao, J.~Ma, et~al.
\newblock Experimental discovery of {Weyl} semimetal {TaAs}.
\newblock {\em Phys. Rev. X}, 5(3):031013, 2015.

\bibitem{xu2015discovery}
S.~Y. Xu, I.~Belopolski, N.~Alidoust, M.~Neupane, G.~Bian, C.~L. Zhang, et~al.
\newblock Discovery of a {Weyl} fermion semimetal and topological {Fermi} arcs.
\newblock {\em Science}, 349(6248):613--617, 2015.

\bibitem{weng2015quantum}
H.~M. Weng, R.~Yu, X.~Hu, X.~Dai, and Z.~Fang.
\newblock Quantum anomalous {Hall} effect and related topological electronic states.
\newblock {\em Advances in Physics}, 64(3):227--282, 2015.

\bibitem{jia2016weyl}
S.~Jia, S.~Y. Xu, and M.~Z. Hasan.
\newblock {Weyl} semimetals, {Fermi} arcs and chiral anomalies.
\newblock {\em Nature materials}, 15(11):1140--1144, 2016.

\bibitem{Colladay:1998fq}
D.~Colladay and V.~A. Kostelecky.
\newblock {Lorentz violating extension of the standard model}.
\newblock {\em Phys. Rev. D}, 58:116002, 1998.

\bibitem{Lai2018WeylKondo}
H.~H. Lai, S.~E. Grefe, S.~Paschen, and Q.~Si.
\newblock {Weyl--Kondo} semimetal in heavy-fermion systems.
\newblock {\em Proc. Nat. Acad. Sci}, 115:93--97, 2018.

\bibitem{Dzsaber:2021ucs}
S.~Dzsaber, X.~Yan, M.~Taupin, G.~Eguchi, Prokofiev A., et~al.
\newblock {Giant spontaneous Hall effect in a nonmagnetic Weyl{\textendash}Kondo semimetal}.
\newblock {\em Proc. Nat. Acad. Sci.}, 118(8):e2013386118, 2021.

\bibitem{Maldacena:1997re}
J.~M. Maldacena.
\newblock {The Large $N$ limit of superconformal field theories and supergravity}.
\newblock {\em Adv. Theor. Math. Phys.}, 2:231--252, 1998.

\bibitem{Gursoy:2012ie}
U.~Gursoy, V.~Jacobs, E.~Plauschinn, H.~Stoof, and S.~Vandoren.
\newblock {Holographic models for undoped Weyl semimetals}.
\newblock {\em JHEP}, 04:127, 2013.

\bibitem{Landsteiner:2015lsa}
K.~Landsteiner and Y.~Liu.
\newblock {The holographic Weyl semi-metal}.
\newblock {\em Phys. Lett. B}, 753:453--457, 2016.

\bibitem{BitaghsirFadafan:2020lkh}
K.~Bitaghsir~Fadafan, A.~O'Bannon, R.~Rodgers, and M.~Russell.
\newblock {A Weyl semimetal from AdS/CFT with flavour}.
\newblock {\em JHEP}, 04:162, 2021.

\bibitem{Furukawa:2024zet}
H.~Furukawa, S.~Ployet, and R.~Rodgers.
\newblock {Conductivities and excitations of a holographic flavour brane Weyl semimetal}.
\newblock {\em JHEP}, 07:115, 2025.

\bibitem{Landsteiner:2015pdh}
K.~Landsteiner, Y.~Liu, and Y.~W. Sun.
\newblock {Quantum phase transition between a topological and a trivial semimetal from holography}.
\newblock {\em Phys. Rev. Lett.}, 116(8):081602, 2016.

\bibitem{Liu:2018djq}
Y.~Liu and Y.~W. Sun.
\newblock {Topological invariants for holographic semimetals}.
\newblock {\em JHEP}, 10:189, 2018.

\bibitem{Chen:2025akz}
X.~T. Chen, X.T. Ji, and Y.~W. Sun.
\newblock {Topological invariant for holographic Weyl-Z$_{2}$ semimetal}.
\newblock {\em JHEP}, 08:048, 2025.

\bibitem{Ammon:2016mwa}
M.~Ammon, M.~Heinrich, A.~Jim{\'e}nez-Alba, and S.~Moeckel.
\newblock {Surface States in Holographic Weyl Semimetals}.
\newblock {\em Phys. Rev. Lett.}, 118(20):201601, 2017.

\bibitem{Juricic:2024tbe}
V.~Juri{\v{c}}i{\'c}, O.~Miskovic, and F.~R. Carrasco.
\newblock {Holographic Weyl semimetals with dislocations}.
\newblock 10 2024.

\bibitem{Juricic:2020sgg}
V.~Juri{\v{c}}i{\'c}, I.~S. Landea, and R.~Soto-Garrido.
\newblock {Phase transitions in a holographic multi-Weyl semimetal}.
\newblock {\em JHEP}, 07:052, 2020.

\bibitem{Liu:2020ymx}
Y.~Liu and X.~M. Wu.
\newblock {An improved holographic nodal line semimetal}.
\newblock {\em JHEP}, 05:141, 2021.

\bibitem{Ahn:2024ozz}
Y.~Ahn, M.~Baggioli, Y.~Liu, and X.M. Wu.
\newblock {Chiral magnetic waves in strongly coupled Weyl semimetals}.
\newblock {\em JHEP}, 03:124, 2024.

\bibitem{Seo:2025pnl}
J.W. Seo, T.~Yuk, and S.J. Sin.
\newblock {Tilted Dirac cones and their topology in Holographic Materials}.
\newblock {\em arXiv: 2509.03033}, 9 2025.

\bibitem{Karch:2002sh}
A.~Karch and E.~Katz.
\newblock {Adding flavor to AdS / CFT}.
\newblock {\em JHEP}, 06:043, 2002.

\bibitem{Iqbal:2009fd}
N.~Iqbal and H.~Liu.
\newblock {Real-time response in AdS/CFT with application to spinors}.
\newblock {\em Fortsch. Phys.}, 57:367--384, 2009.

\bibitem{Liu:2009dm}
H.~Liu, J.~McGreevy, and D.~Vegh.
\newblock {Non-Fermi liquids from holography}.
\newblock {\em Phys. Rev. D}, 83:065029, 2011.

\bibitem{Lu:2024qxj}
C.~Y. Lu, X.~H. Ge, and S.~J. Sin.
\newblock {Holographic fermions in the dyonic Gubser-Rocha black hole}.
\newblock {\em Phys. Rev. D}, 111(8):086011, 2025.

\bibitem{Fang:2014jka}
L.~Q. Fang, X.~H. Ge, J.~P. Wu, and H.~Q. Leng.
\newblock {Anisotropic Fermi surface from holography}.
\newblock {\em Phys. Rev. D}, 91(12):126009, 2015.

\bibitem{Fang:2012pw}
L.~Q. Fang, X.~H. Ge, and X.~M. Kuang.
\newblock {Holographic fermions in charged Lifshitz theory}.
\newblock {\em Phys. Rev. D}, 86:105037, 2012.

\bibitem{Fang:2013ixa}
L.~Q. Fang, X.~H. Ge, and Xi.~M. Kuang.
\newblock {Holographic fermions with running chemical potential and dipole coupling}.
\newblock {\em Nucl. Phys. B}, 877:807--824, 2013.

\bibitem{Ge:2023ghy}
X.~H. Ge, S.~Ishigaki, S.~J. Sin, and T.~Yuk.
\newblock {First order phase transition in the D3-D7 model from the point of view of the fermionic spectral functions}.
\newblock {\em Phys. Rev. D}, 110(2):026003, 2024.

\bibitem{Ammon:2010pg}
M.~Ammon, J.~Erdmenger, M.~Kaminski, and A.~O'Bannon.
\newblock {Fermionic Operator Mixing in Holographic p-wave Superfluids}.
\newblock {\em JHEP}, 05:053, 2010.

\bibitem{Kirsch:2006he}
I.~Kirsch.
\newblock {Spectroscopy of fermionic operators in AdS/CFT}.
\newblock {\em JHEP}, 09:052, 2006.

\bibitem{Abt:2019tas}
R.~Abt, J.~Erdmenger, N.~Evans, and K.~S. Rigatos.
\newblock {Light composite fermions from holography}.
\newblock {\em JHEP}, 11:160, 2019.

\bibitem{lifshitz1960anomalies}
I.~M. Lifshitz.
\newblock Anomalies of electron characteristics of a metal in the high pressure region.
\newblock {\em Sov. Phys. JETP}, 11(5):1130--1135, 1960.

\bibitem{polchinskistring}
Joseph Polchinski.
\newblock String theory. vol. 2: superstring theory and beyond 1998.
\newblock {\em Univ. Pr.. Cambridge, UK: r. 531p}.

\bibitem{Hashimoto:2013mua}
K.~Hashimoto and T.~Oka.
\newblock {Vacuum Instability in Electric Fields via AdS/CFT: Euler-Heisenberg Lagrangian and Planckian Thermalization}.
\newblock {\em JHEP}, 10:116, 2013.

\bibitem{Hashimoto:2014yya}
K.~Hashimoto, T.~Oka, and A.~Sonoda.
\newblock {Electromagnetic instability in holographic QCD}.
\newblock {\em JHEP}, 06:001, 2015.

\bibitem{Faulkner:2009wj}
T.~Faulkner, H.~Liu, J.~McGreevy, and D.~Vegh.
\newblock {Emergent quantum criticality, Fermi surfaces, and AdS(2)}.
\newblock {\em Phys. Rev. D}, 83:125002, 2011.

\bibitem{Faulkner:2011tm}
T.~Faulkner, N.~Iqbal, H.~Liu, J.~McGreevy, and D.~Vegh.
\newblock {Holographic non-Fermi liquid fixed points}.
\newblock {\em Phil. Trans. Roy. Soc.}, A 369:1640, 2011.

\bibitem{Fang:2015dia}
L.~Q. Fang, X.~M. Kuang, B.~Wang, and J.~P. Wu.
\newblock {Fermionic phase transition induced by the effective impurity in holography}.
\newblock {\em JHEP}, 11:134, 2015.

\bibitem{Mateos:2007vn}
D.~Mateos, R.~C.. Myers, and R.~M. Thomson.
\newblock {Thermodynamics of the brane}.
\newblock {\em JHEP}, 05:067, 2007.

\bibitem{Albash:2006bs}
T.~Albash, V.~G. Filev, C.~V. Johnson, and A.~Kundu.
\newblock {Global Currents, Phase Transitions, and Chiral Symmetry Breaking in Large N(c) Gauge Theory}.
\newblock {\em JHEP}, 12:033, 2008.

\bibitem{Kruczenski:2003be}
M.~Kruczenski, D.~Mateos, R.~C. Myers, and D.~J. Winters.
\newblock {Meson spectroscopy in AdS / CFT with flavor}.
\newblock {\em JHEP}, 07:049, 2003.

\bibitem{Karch:2007pd}
A.~Karch and A.~O'Bannon.
\newblock {Metallic AdS/CFT}.
\newblock {\em JHEP}, 09:024, 2007.

\bibitem{Plantz:2018tqf}
N.~W.~M. Plantz, F.~Garc{\'\i}a~Fl{\'o}rez, and H.~T.~C. Stoof.
\newblock {Massive Dirac fermions from holography}.
\newblock {\em JHEP}, 04:123, 2018.

\bibitem{Yuk:2022lof}
T.~Yuk and S.~J. Sin.
\newblock {Flow equation and fermion gap in the holographic superconductors}.
\newblock {\em JHEP}, 02:121, 2023.

\bibitem{Volovik:2016dsj}
G.~E. Volovik.
\newblock {Topological Lifshitz transitions}.
\newblock {\em Low Temp. Phys.}, 43(1):47, 2017.

\bibitem{liu2010evidence}
C.~Liu, T.~Kondo, R.~M. Fernandes, A.~D. Palczewski, E.~D. Mun, et~al.
\newblock Evidence for a lifshitz transition in electron-doped iron arsenic superconductors at the onset of superconductivity.
\newblock {\em Nature Physics}, 6(6):419--423, 2010.

\bibitem{ren2017superconductivity}
M.Q. Ren, Y.~J. Yan, X.~H. Niu, R.~Tao, D.~Hu, et~al.
\newblock Superconductivity across lifshitz transition and anomalous insulating state in surface {K}--dosed ({Li}$_{0.8}${Fe}$_{0.2}${OH}){FeSe}.
\newblock {\em Science Advances}, 3(7):e1603238, 2017.

\bibitem{xu2018evidence}
N~Xu, Z.~W. Wang, A.~Magrez, P.~Bugnon, H.~Berger, et~al.
\newblock Evidence of a coulomb-interaction-induced {Lifshitz} transition and robust hybrid {Weyl} semimetal in ${T_d}$-{MoTe}$_2$.
\newblock {\em Phys. Rev. Lett.}, 121(13):136401, 2018.

\bibitem{lohani2023electronic}
H.~Lohani, P.~Foulquier, P.~Le~Fevre, F.~Bertran, D.~Colson, et~al.
\newblock Electronic structure evolution of the magnetic {Weyl} semimetal {Co}$_3${Sn}$_2$s$_2$ with hole and electron doping.
\newblock {\em Phys. Rev. B}, 107(24):245119, 2023.

\bibitem{jung2024quantum}
H.~Jung, K.~Jin, M.~Sung, J.~Kim, J.~Kim, et~al.
\newblock Quantum-confined lifshitz transition on weyl semimetal ${T_d}$-{MoTe}$_2$.
\newblock {\em ACS nano}, 18(34):23189--23195, 2024.

\bibitem{lin2010lifshitz}
J.~Lin.
\newblock Lifshitz transition in two-dimensional spin-density wave models.
\newblock {\em Phys. Rev. B}, 82(19):195110, 2010.

\bibitem{chen2012lifshitz}
K.~S. Chen, Z.~Y. Meng, T.~Pruschke, J.~Moreno, and M.~Jarrell.
\newblock Lifshitz transition in the two-dimensional hubbard model.
\newblock {\em Phys. Rev. B}, 86(16):165136, 2012.

\bibitem{carr2010lifshitz}
S.~T. Carr, J.~Quintanilla, and J.~J. Betouras.
\newblock Lifshitz transitions and crystallization of fully polarized dipolar fermions in an anisotropic two-dimensional lattice.
\newblock {\em Phys. Re. B}, 82(4):045110, 2010.

\bibitem{yang2019topological}
H.~F. Yang, L.~X. Yang, Z.~K. Liu, Y.~Sun, C.~Chen, et~al.
\newblock Topological {Lifshitz} transitions and {Fermi} arc manipulation in {Weyl} semimetal {NbAs}.
\newblock {\em Nature communications}, 10(1):3478, 2019.

\bibitem{bragancca2018correlation}
H.~Bragan{\c{c}}a, S.~Sakai, MCO. Aguiar, and M.~Civelli.
\newblock Correlation-driven lifshitz transition at the emergence of the pseudogap phase in the two-dimensional hubbard model.
\newblock {\em Phys. Rev. Lett.}, 120(6):067002, 2018.

\end{thebibliography}


\end{document}